\documentclass{iopart}
\usepackage{epsfig}
\begin{document}

\title[The $\nu_e \to \nu_\tau$ channel 
as a tool to solve ambiguities]{The $\nu_e \to \nu_\tau$ channel 
as a tool to solve ambiguities\footnote{Talk presented by Davide Meloni.}}

\author{A. Donini\dag, D. Meloni\dag~and P. Migliozzi$^\star$}

\address{\dag\  Dip. Fisica, Universit\`a di Roma ``La Sapienza''
and I.N.F.N., Sezione di Roma I, P.le A. Moro 2, I-00185, Rome, Italy }

\address{$\star$ I.N.F.N., Sezione di Napoli, 
Complesso Universitario di Monte Sant'Angelo, 
Via Cintia ed. G, I-80126 Naples, Italy }

\begin{abstract}
In this talk I show how considering at the same time the 
$\nu_e \to \nu_\tau$ and the $\nu_e \to \nu_\mu$ oscillation channels
the errors in the 
leptonic CP-violating phase $\delta$ measurement could significantly be reduced.

\end{abstract}

The present atmospheric \cite{Toshito:2001dk} 
and solar \cite{Fukuda:1996sz} neutrino data 
are strongly supporting the hypothesis of neutrino oscillations and can be 
easily accommodated in a three family mixing scenario. In particular, data on 
atmospheric neutrinos are interpreted as oscillations 
of muon neutrinos into neutrinos that are not $\nu_e$'s, with  
the corresponding mixing angle  
$\sin^2 2 \theta_{23} > 0.92$ and $|\Delta m^2_{23}|$ in the range 
$1.5$ to $3.8 \times 10^{-3}$ eV$^2$ \cite{Nakaya:2002ki}.
The recent SNO results for solar neutrinos \cite{Ahmad:2001an}
favour the LMA-MSW solution of the solar neutrino deficit with 
$\nu_e$ oscillations into active neutrino states with a large corresponding
mixing angle ($\theta_{12}$). 

By the end of the current decade, the $\theta_{13}$ angle
could still be poorly known (or even unknown) and no information whatsoever
will be at hand regarding the leptonic CP violating phase $\delta$. For this reason it has been
proposed to build the ``Neutrino Factory'' \cite{Geer:1998iz,DeRujula:1998hd} 
and suitably optimized detectors located far away from the neutrino source. 

\section*{The problem...}

In \cite{Burguet-Castell:2001ez} it has been noticed that the probability 
$P_{\alpha \beta} (\bar \theta_{13},\bar \delta)$ obtained 
for neutrinos at a fixed energy and baseline with input parameter 
($\bar \theta_{13},\bar \delta$), can be reproduced
varying accordingly the values of $\theta_{13}$ and $\delta$. 
Considering at the same time the equiprobability curve for antineutrinos 
at the same energy and with the same input parameters, the two equiprobability 
curves have two intersections: 
the input pair ($\bar \theta_{13},\bar \delta$) and 
a second, energy dependent, point ($\tilde \theta_{13},\tilde \delta$), the
"clone".
This second intersection introduces an ambiguity in the measurement 
of the physical values of $\theta_{13}$ and $\delta$ (the so-called
($\theta_{13}, \delta$) ambiguity). In this case a fitting
procedure to reconstruct the physical parameters will identify two 
low $\chi^2$ regions: one close to the input value and the other in the
restricted area in which the other intersections are present (a
"clone" region).

\section*{...and its solution}
\begin{figure}[h!]
\begin{center}
\epsfxsize6.5cm\epsffile{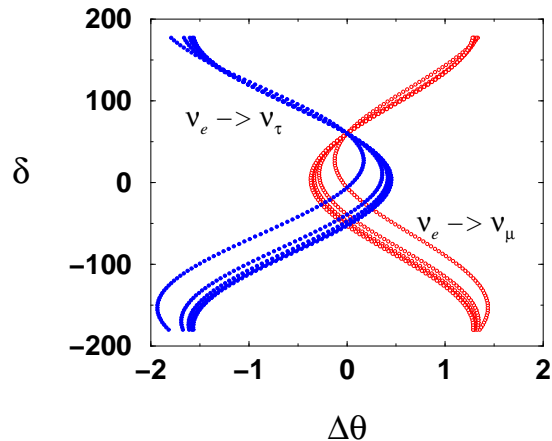} 
\caption{\it Superposition of the equal-number-of-events curves for the transition 
$\nu_e \to \nu_\mu$ (light lines) and $\nu_e \to \nu_\tau$ (dark lines), 
for $L = 732$ Km and $\bar \theta_{13} = 5^\circ, \bar \delta = 60^\circ$.
$\Delta \theta$ is the shift of $\theta_{13}$ with respect to its input value 
$\bar \theta_{13}$. Four ENE curves for each transition have been shown
corresponding to four neutrino energy bins, from 10 GeV to 50 GeV.}
\label{fig:mudistr}
\end{center}
\end{figure}

Different proposals have been suggested
to solve this ambiguity (see \cite{Burguet-Castell:2001ez,Freund:2001ui} for 
possible solutions to the problem using the combination of different baselines
or detectors with improved energy resolution).
We notice \cite{Donini:2002rm} that muons proceeding from 
$\tau$ decay when $\tau$'s are produced via a $\nu_e \to \nu_\tau$ transition
(the "silver channel") show a 
different $(\theta_{13}, \delta)$ correlation from those coming from 
$\nu_e \to \nu_\mu$ (the "golden channel"). This can be seen by 
looking at the 
equal-number-of-events (ENE) curves, computed solving $N^i(\theta_{13}, \delta) = 
N^i (\bar \theta_{13}, \bar \delta)$ , 
where $N^i$ is the number of "wrong-sign" muons produced in the
detector in the i-th energy bin (see Fig.\ref{fig:mudistr} and 
\cite{Donini:2002rm} for details). 

The different ($\theta_{13}, \delta$) correlation between two channels comes 
from the different structure of the transition probabilities, which have an 
opposite sign
in front of the CP-violating term. In principle, this different behaviour of the ENE curves should
reduce or eliminate the impact of the clone solutions when fitting
simultaneously the two sets of data to reconstruct the physical parameters.
To take full advantage from the ``silver'' channel, we should use a detector able 
to distinguish muons originated from $\tau$ decay from the 
``golden'' muons.  For this reason we consider an OPERA-like detector with a
mass of 2 Kton and spectrometers capable of muon charge identification
(see the OPERA proposal for details, \cite{Guler:2000bd}) located at
$L = 732$ Km down the neutrino source. 

\begin{figure}[h!]
\begin{center}
\begin{tabular}{cc}
\epsfxsize5.7cm\epsffile{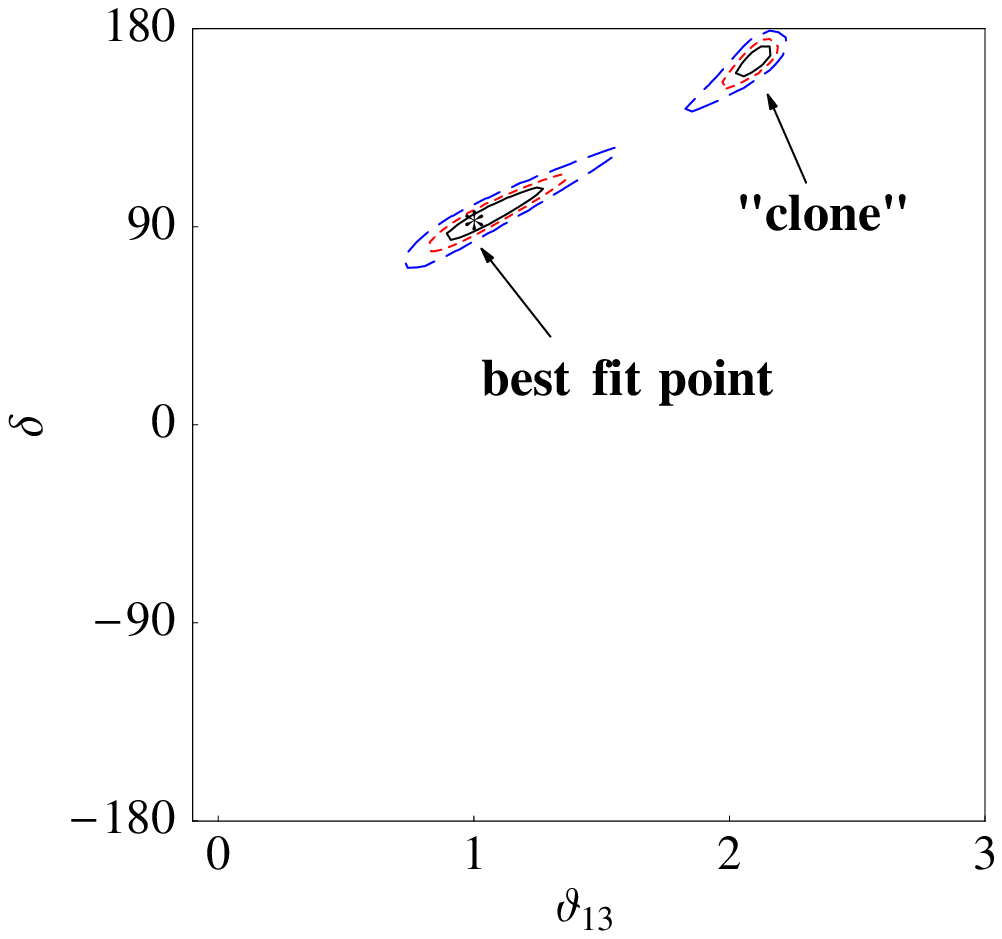} &
              \epsfxsize5.7cm\epsffile{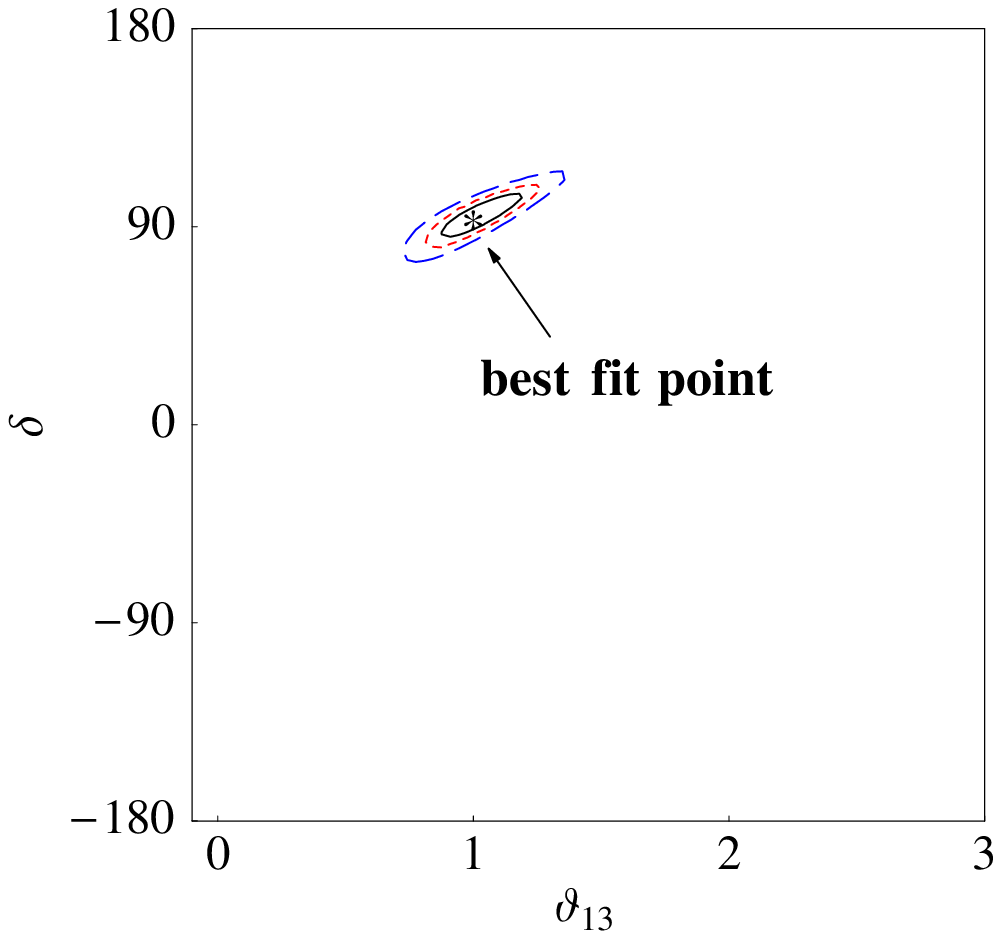}
\end{tabular}

\caption{ \it 68.5, 90 and 99 \% C.L. contours resulting from a 
$\chi^2$ fit of $\theta_{13}$
and $\delta$, for $\bar \theta_{13} = 1^\circ$ and 
$\bar \delta = 90^\circ$, for the combination of an iron 
detector at $L = 3000$ Km and an emulsion detector at $L = 732$ Km:
(left) only ``golden'' muon events are taken into account;
(right) both ``silver'' and ``golden'' muon events are taken into account. 
Five years of data taking for both polarities 
in the distant detector and only five years in the $\mu^+$ polarity in the near detector
have been considered.}
\label{fig:notau}
\end{center}
\end{figure}

The improvement in the
reconstruction of the physical parameters can be seen comparing the plots 
in Fig.\ref{fig:notau}. On the left we consider two realistic iron detector (including efficiencies and 
background as quoted in \cite{Cervera:2000kp}) located at two different 
baselines $L = 732$ Km and $L = 3000$ Km (the 
optimal distance for the measurement of leptonic CP violation). The best fit point, obtained
following the strategy presented in \cite{Cervera:2000kp}, is close to the
input parameter ($\bar \theta_{13} = 1^\circ$ and $\bar \delta = 90^\circ$) but
a clone region is present and the determination of the physical parameters
is affected by ambiguity. On the right plot, we show the results of the fit 
obtained combining at the same time both the
``golden'' and the ``silver'' channels at the ideal near emulsion detector 
and the ``golden'' channel at the magnetized iron detector 
located at $L = 3000$ 
Km. The clone region completely disappears and the input parameters can be
determined with an error of tens of degrees on $\delta$ and tenths of degrees on 
$\theta_{13}$.

Considering a realistic estimate of the reconstruction
efficiency and of the main backgrounds both for ``golden'' and
``silver'' muon events at the emulsion detector \cite{Guler:2000bd}, 
the clone regions for $\theta_{13} \simeq 1^\circ$ do appear and our results
 are comparable to those obtained with a combination of two realistic
magnetized iron detectors located at $L = 732$ Km and $L = 3000$ Km. 
However, at the time the Neutrino Factory will be operational, 
several improvement of the lead-emulsion detector could be done. In particular, we observed that 
a moderate scaling in the detector mass from 2 to 4 Kton completely eliminates 
the clone regions for any value of $\theta_{13} \geq 1^\circ$. On the other
hand, a moderate increase in the main background rejection
does not seem to improve in a significant way the previous results.

\end{document}